\documentclass[runningheads]{svmult}

\usepackage{makeidx}   
\usepackage{graphicx}  
\usepackage{subeqnar}  
\usepackage{multicol}  
\usepackage{physmubb}  
\makeindex             

\usepackage{epsfig}
\newcommand{\agt}{\,\rlap{\lower 3.5 pt \hbox{$\mathchar \sim$}} \raise 1pt
 \hbox {$>$}\,}
\newcommand{\alt}{\,\rlap{\lower 3.5 pt \hbox{$\mathchar \sim$}} \raise 1pt
 \hbox {$<$}\,}

\begin{document}

\title*{Hadron Production in Hadron-Hadron and Lepton-Hadron Collisions%
\thanks{To appear in the {\it Proceedings of the 14th Topical Conference on
Hadron Collider Physics (HCP 2002)}, 29 September -- 4 October 2002,
Karlsruhe, Germany.}
}

\toctitle{Hadron Production in Hadron-Hadron and
\protect\newline Lepton-Hadron Collisions}

\titlerunning{Hadron Production}

\author{Bernd A. Kniehl\\
II. Institut f\"ur Theoretische Physik, Universit\"at Hamburg,
Luruper Chaussee 149, 22761 Hamburg, Germany}

\authorrunning{Bernd A. Kniehl}

\maketitle              

\begin{abstract}
We review a recent global analysis of inclusive single-charged-hadron
production in high-energy colliding-beam experiments, which is performed at
next-to-leading order (NLO) in the parton model of quantum chromodynamics
endowed with nonperturbative fragmentation functions (FFs).
Comparisons of $p\overline{p}$ data from CERN S$p\overline{p}$S and the
Fermilab Tevatron and $\gamma p$ data from DESY HERA with the corresponding
NLO predictions allow for quantitative tests of the scaling violations in the
FFs and their universality.
We emphasize the potential of new measurements at the Tevatron to place tight
constraints in the large-$x$ region and on the gluon FF, complementary to
those from $e^+e^-$ data, which are indispensible in order to reliably predict
the $\pi^0$ background for the $H\to\gamma\gamma$ signal of the
intermediate-mass Higgs boson at the Tevatron and the CERN LHC.
Adopting a similar theoretical framework for $b$-hadron production, we show
that the notorious excess of the Tevatron data over existing theoretical
calculations can be ascribed, at sufficiently large values of $p_T$, to
nonperturbative fragmentation effects inadequately included previously.
\end{abstract}

\section{Introduction}

In the framework of the QCD-improved parton model, the inclusive production of
single hadrons is described by means of fragmentation functions (FFs)
$D_a^h(x,\mu^2)$.
The value of $D_a^h(x,\mu^2)$ corresponds to the probability for the parton
$a$ produced at short distance $1/\mu$ to form a jet that includes the hadron
$h$ carrying the fraction $x$ of the longitudinal momentum of $a$.
Unfortunately, it is not yet possible to calculate the FFs from first
principles, in particular for hadrons with masses smaller than or comparable
to the asymptotic scale parameter $\Lambda$.
However, given their $x$ dependence at some energy scale $\mu$, the evolution
with $\mu$ may be computed perturbatively in QCD using the timelike 
Altarelli-Parisi (AP) equations \cite{gri}.
This allows us to test QCD quantitatively within one experiment observing
single hadrons at different values of centre-of-mass (CM) energy $\sqrt s$ (in
the case of $e^+e^-$ annihilation) or transverse momentum $p_T$ (in the case
of scattering).
Moreover, the factorization theorem guarantees that the $D_a^h(x,\mu^2)$
functions are independent of the process in which they have been determined
and represent a universal property of $h$.
This enables us to make quantitative predictions for other types of
experiments as well.

During the last seven years, the experiments at CERN LEP1 and SLAC SLC have
provided us with a wealth of high-precision information on how partons
fragment into low-mass charged hadrons ($h^\pm$).
The data partly comes as light-, $c$-, and $b$-quark-enriched samples without
\cite{Al,A} or with identified final-state hadrons ($\pi^\pm$, $K^\pm$, and
$p/\overline{p}$) \cite{A1} or as gluon-tagged three-jet samples without
hadron identification \cite{Ag,Dg}.
Motivated by this new situation, the author, together with Kramer and
P\"otter, recently updated, refined, and extended a previous analysis
\cite{bkk} by generating new leading-order (LO) and next-to-leading-order
(NLO) sets of $\pi^\pm$, $K^\pm$, and $p/\overline{p}$ FFs \cite{kkp}.
By also including in our fits $\pi^\pm$, $K^\pm$, and $p/\overline{p}$ data
(without flavour separation) from PEP \cite{T}, with CM energy
$\sqrt s=29$~GeV, we obtained a handle on the scaling violations in the FFs,
which enabled us to determine the strong-coupling constant
$\alpha_s^{(5)}(M_Z)$.

The formation of $D$ and $B$ mesons from $c$ and $b$ quarks, respectively, is
a genuinely nonperturbative process, so that it is indispensable to employ
nonperturbative FFs.
Furthermore, if the characteristic energy scale $\mu$ is large against the
heavy-quark mass $m_c$ ($m_b$), then the QCD-improved parton model with
$n_f=4$ ($n_f=5$) active quark flavours is the appropriate theoretical
framework, which allows for coherent analyses of data from $e^+e^-$,
lepton-hadron, and hadron-hadron collisions \cite{dst,bme}.

This contribution is organized as follows.
In Sect.~\ref{sec:two}, we present some details of our global fits \cite{kkp}
and assess the quality of the resulting FFs.
We also discuss the determination of $\alpha_s^{(5)}(M_Z)$ from the scaling
violations in the FFs \cite{kkp1}.
In Sect.~\ref{sec:three}, we present comparisons of our NLO predictions for
inclusive charged-hadron production \cite{kkp2} with $p\overline{p}$ data from
S$p\overline{p}$S \cite{UA1} and the Tevatron \cite{CDF} and with $\gamma p$
data from HERA \cite{H1}.
In Sect.~\ref{sec:four}, we demonstrate that the $p_T$ distribution of
inclusive $B$-meson hadroproduction measured by CDF \cite{cdf1,cdf2} is well
described at NLO in the QCD parton model with $n_f=5$ endowed with
nonperturbative FFs, once the latter are fitted to LEP1 data \cite{ob}.
Our conclusions are summarized in Sect.~\ref{sec:five}.

\section{Determination of the FFs
\label{sec:two}}

The NLO formalism for extracting FFs from measurements of the cross section
$d\sigma/dx$ of $e^+e^-\to h+X$ is comprehensively described in \cite{bkk}.
We work in the modified minimal-subtraction ($\overline{\mathrm{MS}}$)
renormalization and factorization scheme and choose the renormalization and
factorization scales to be $\mu_R=\mu_F=\xi\sqrt s$, except for gluon-tagged
three-jet events, where we put $\mu_R=\mu_F=\xi\times2E_{\mathrm{jet}}$, with
$E_{\mathrm{jet}}$ being the gluon-jet energy in the CM frame.
Here, the dimensionless parameter $\xi$ is introduced to determine the
theoretical uncertainty in $\alpha_s^{(5)}(M_Z)$ from scale variations.
As usual, we allow for variations of $\xi$ between $1/2$ and 2 about the
default value 1.

\begin{figure}[t]
\begin{center}
\begin{tabular}{ll}
\parbox{5.65cm}{
\epsfig{file=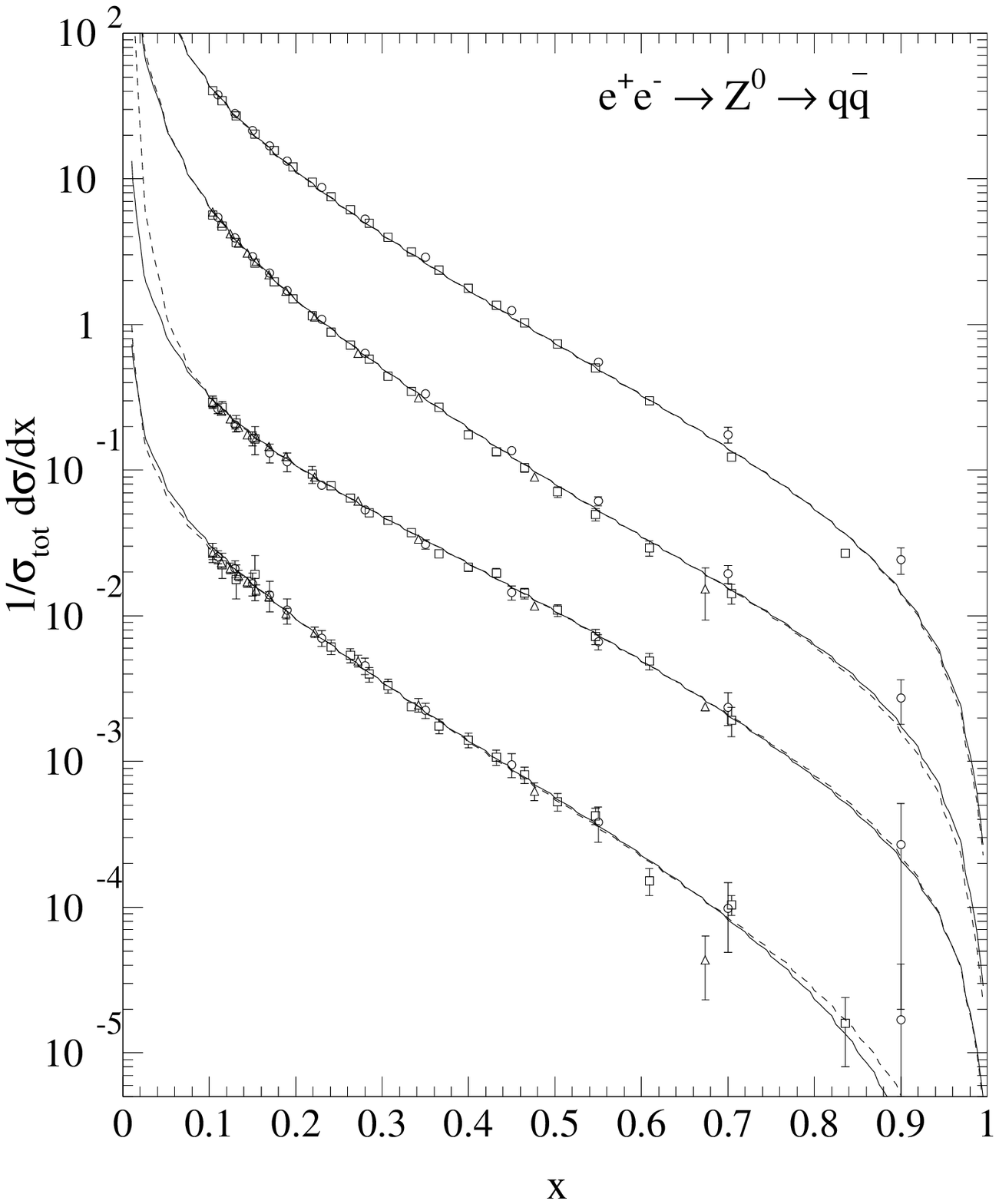,width=5.65cm,bbllx=4pt,bblly=11pt,bburx=428pt,%
bbury=526pt,clip=}
} &
\parbox{5.65cm}{
\epsfig{file=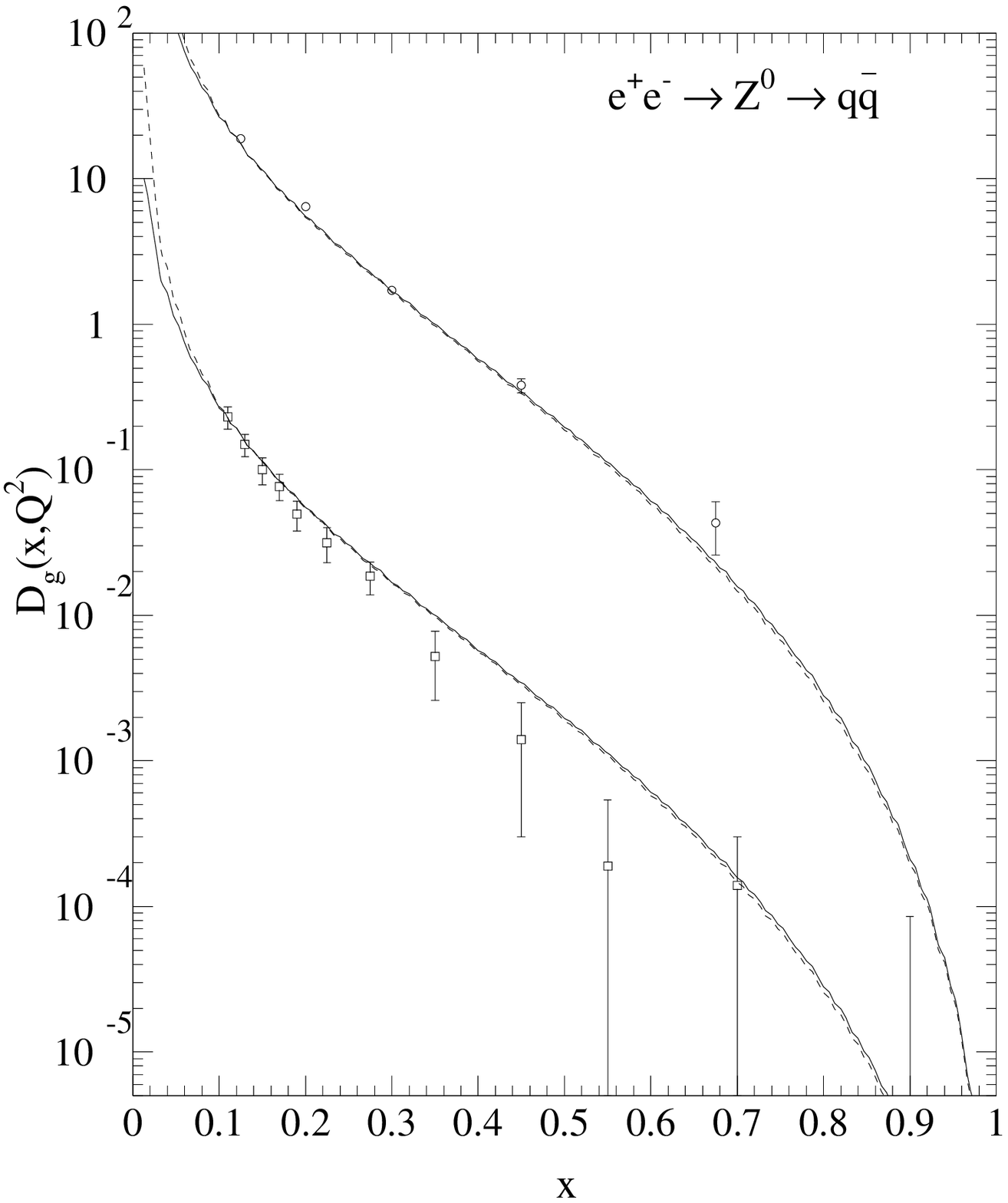,width=5.65cm,bbllx=0pt,bblly=10pt,bburx=428pt,%
bbury=526pt,clip=}
}
\vspace*{-0.3cm} \\
(a) & (b) \\
\end{tabular}
\caption[]{({\bf a}) Comparison of data on inclusive charged-hadron production
at $\protect\sqrt{s}=91.2$~GeV from ALEPH ({\it triangles}), DELPHI
({\it circles}), and SLD ({\it squares}) \cite{A1} with our LO ({\it dashed
lines}) and NLO ({\it solid lines}) fit results \cite{kkp}.
The upmost, second, third, and lowest curves refer to charged hadrons,
$\pi^\pm$, $K^\pm$, and $p/\overline{p}$, respectively.
({\bf b}) Comparison of three-jet data on the gluon FF from ALEPH with
$E_{\mathrm{jet}}=26.2$~GeV ({\it upper curves}) and from OPAL with
$E_{\mathrm{jet}}=40.1$~GeV ({\it lower curves}) \cite{Ag} with our LO
({\it dashed lines}) and NLO ({\it solid lines}) fit results \cite{kkp}}
\label{fig:fit}
\end{center}
\end{figure}

Our strategy was to only include in our fits LEP1 and SLC data with both
flavour separation and hadron identification \cite{A1}, gluon-tagged
three-jet samples with a fixed gluon-jet energy \cite{Ag}, and the
$\pi^\pm$, $K^\pm$, and $p/\overline{p}$ data sets from the pre-LEP1/SLC era
with the highest statistics and the finest binning in $x$ \cite{T}.
The $\chi^2$ value per fitted data point turned out to be
$\chi^2_{\mathrm{DF}}=0.97$ (0.98) at LO (NLO).
The goodness of our fit may also be judged from Figs.~\ref{fig:fit}(a) and
(b), where our LO and NLO fit results are compared with the ALEPH, DELPHI,
OPAL, and SLD data \cite{A1,Ag}.
Other data served us for cross checks \cite{Al,A,Dg,low,Dl}.
In particular, we probed the scaling violations in the FFs through comparisons
with $\pi^\pm$, $K^\pm$, and $p/\overline{p}$ data from DESY DORIS and PETRA,
with CM energies between 5.4 and 34~GeV \cite{low}.
Furthermore, we tested the gluon FF, which enters the unpolarized cross
section only at NLO, by comparing our predictions for the longitudinal cross
section, where it already enters at LO, with available data \cite{Al,Dl}.
Finally, we directly compared our gluon FF with the one recently measured by
DELPHI in three-jet production with gluon identification as a function of $x$
at various energy scales $\mu$ \cite{Dg}.
All these comparisons led to rather encouraging results.
We also verified that our FFs satisfy reasonably well the momentum sum
rules, which we did not impose as constraints on our fits.

Since we included in our fits high-quality data from two very different
energies, namely 29 and 91.2~GeV, we were sensitive to the running of
$\alpha_s(\mu)$ and, therefore, able to extract values of 
$\Lambda_{\overline{\mathrm{MS}}}^{(5)}$.
We obtained
$\Lambda_{\overline{\mathrm{MS}}}^{(5)}=88{+34\atop-31}{+3\atop-23}$~MeV at LO
and $\Lambda_{\overline{\mathrm{MS}}}^{(5)}=213{+75\atop-73}{+22\atop-29}$~MeV
at NLO, where the first errors are experimental and the second ones are
theoretical.
From the LO and NLO formulas for $\alpha_s^{(n_f)}(\mu)$, we thus obtained
$\alpha_s^{(5)}(M_Z)=0.1181{+0.0058\atop-0.0069}{+0.0006\atop-0.0049}$ (LO)
and $\alpha_s^{(5)}(M_Z)=0.1170{+0.0055\atop-0.0069}{+0.0017\atop-0.0025}$
(NLO), respectively, which is already included in the most recent world
average by the Particle Data Group \cite{pdg}.
As expected, the theoretical error on $\alpha_s^{(5)}(M_Z)$ is significantly
reduced as we pass from LO to NLO.
We observe that our LO and NLO values of $\alpha_s^{(5)}(M_Z)$ are quite
consistent with each other, which indicates that our analysis is 
perturbatively stable.
The fact that the respective values of
$\Lambda_{\overline{\mathrm{MS}}}^{(5)}$ significantly differ is a well-known
feature of the $\overline{\mathrm{MS}}$ definition of
$\alpha_s^{(n_f)}(\mu)$ \cite{cks}.

\section{Global Analysis of Collider Data
\label{sec:three}}

\begin{figure}[ht]
\begin{center}
\begin{tabular}{ll}
\parbox{5.65cm}{
\epsfig{file=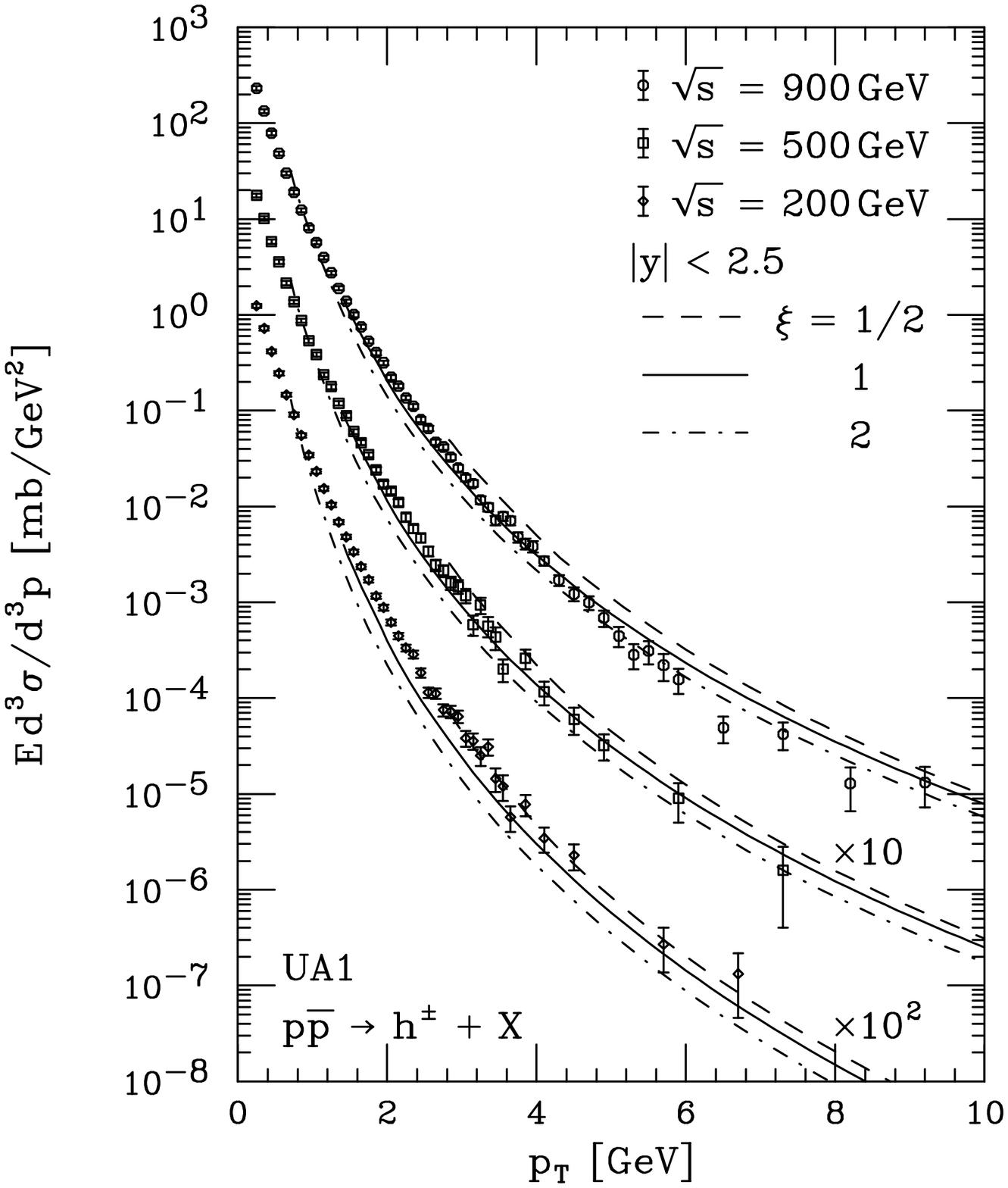,width=5.65cm,bbllx=49pt,bblly=102pt,bburx=542pt,%
bbury=681pt,clip=}
} &
\parbox{5.65cm}{
\epsfig{file=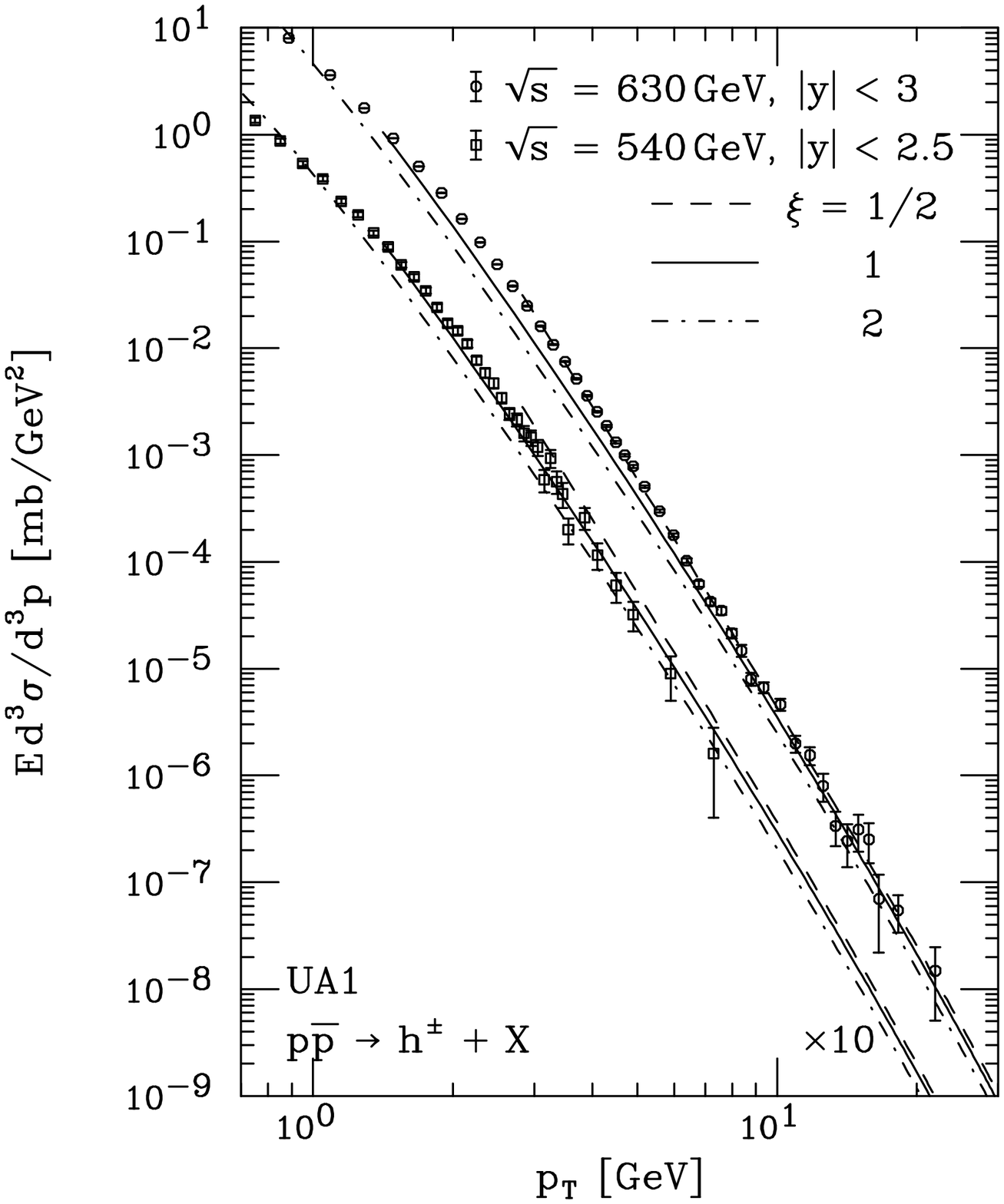,width=5.65cm,bbllx=49pt,bblly=102pt,bburx=542pt,%
bbury=681pt,clip=}
}
\vspace*{-0.3cm} \\
(a) & (b) \\
\parbox{5.65cm}{
\epsfig{file=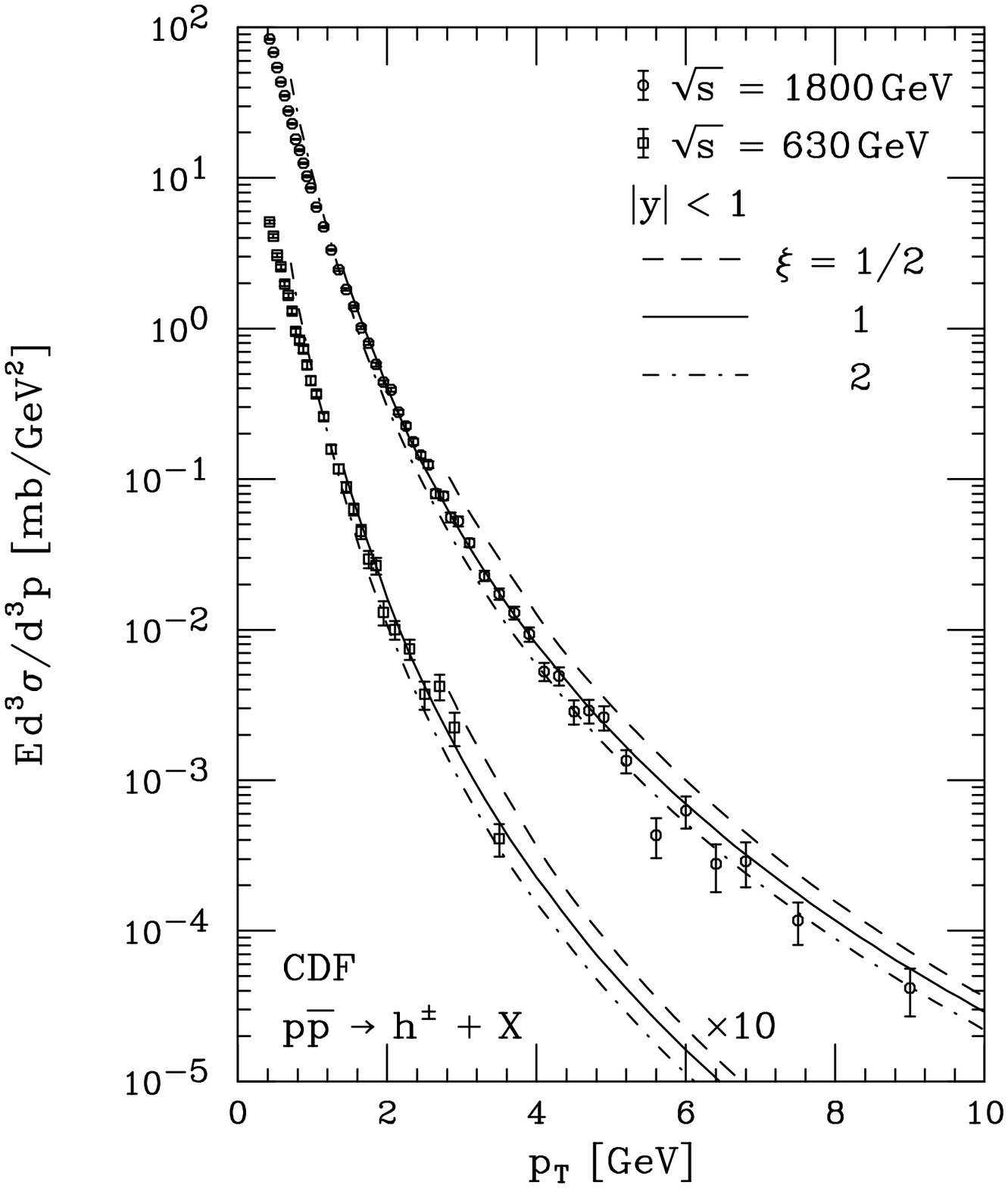,width=5.65cm,bbllx=49pt,bblly=102pt,bburx=542pt,%
bbury=681pt,clip=}
} &
\parbox{5.65cm}{
\epsfig{file=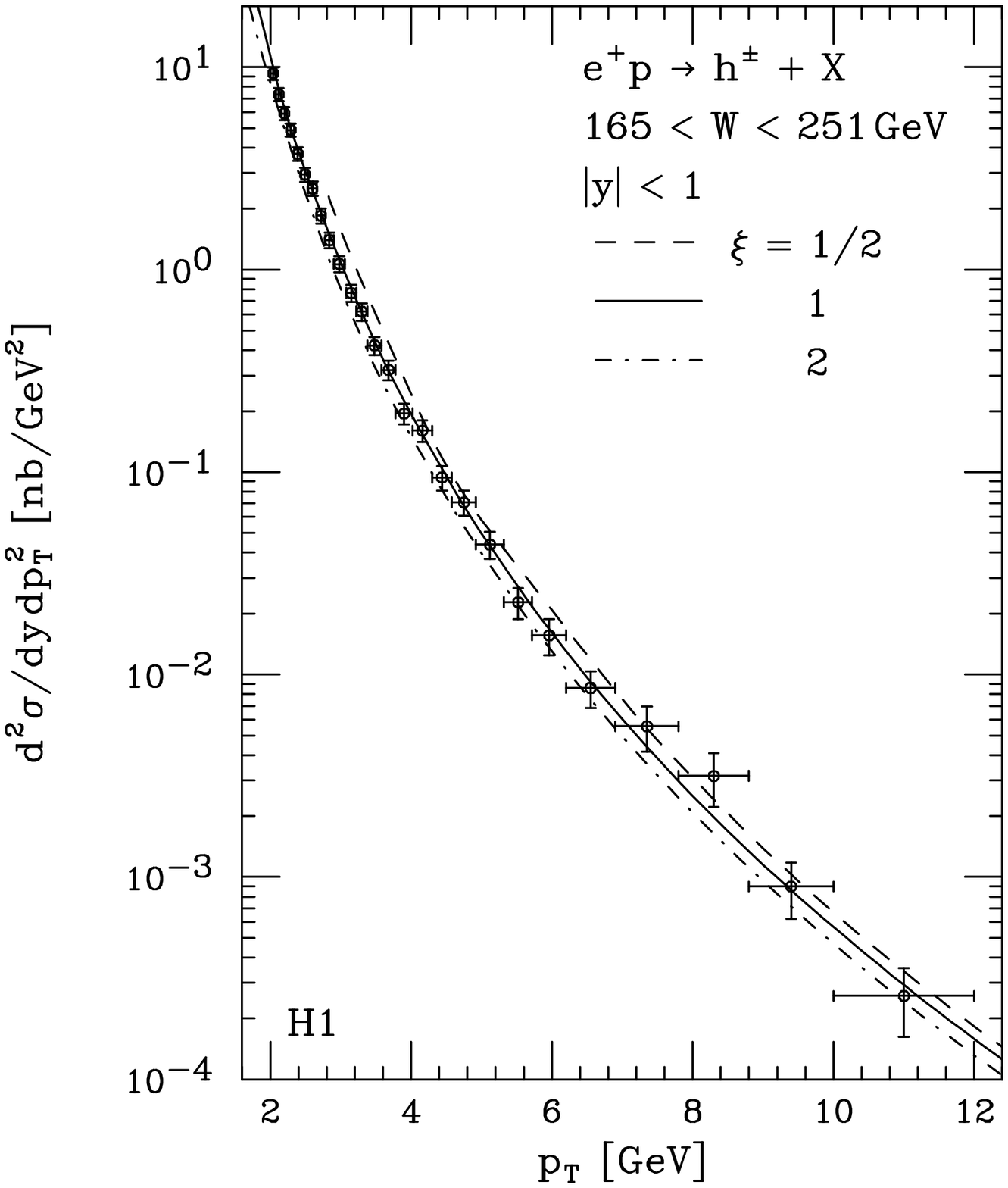,width=5.65cm,bbllx=49pt,bblly=102pt,bburx=532pt,%
bbury=681pt,clip=}
}
\vspace*{-0.3cm} \\
(c) & (d)
\end{tabular}
\caption[]{Comparisons of ({\bf a}, {\bf b}) S$p\overline{p}$S \cite{UA1} and
({\bf c}) Tevatron \cite{CDF} data of $p\overline{p}\to h^\pm+X$ and ({\bf d})
HERA data of $\gamma p\to h^\pm+X$ \cite{H1} with our NLO predictions
\cite{kkp2}}
\label{fig:uni}
\end{center}
\end{figure}

Recently, we extended our previous tests of scaling violations \cite{kkp} to
higher energy scales by confronting new data of $e^+e^-\to h^\pm+X$ from LEP2
\cite{D2}, with $\sqrt s$ ranging from 133~GeV up to 189~GeV, with NLO
predictions based on our FFs \cite{kkp2}.
Furthermore, we quantitatively checked the universality of our FFs by making
comparisons with essentially all available high-statistics data on inclusive
charged-hadron production in colliding-beam experiments \cite{kkp2}.
This includes $p\overline{p}$ data from the UA1 and UA2 Collaborations
\cite{UA1} at S$p\overline{p}$S and from the CDF Collaboration \cite{CDF} at
the Tevatron, $\gamma p$ data from the H1 and ZEUS Collaborations \cite{H1} at
HERA, and $\gamma\gamma$ data from the OPAL Collaboration \cite{Ogg} at LEP2.
In hadroproduction and photoproduction, we set $\mu_R=\mu_F=\xi p_T$.
As for the parton density functions (PDFs) of the proton, we employed set
CTEQ5M provided by the CTEQ Collaboration \cite{CTEQ5}, with
$\Lambda_{\overline{\mathrm{MS}}}^{(5)}=226$~MeV.
As for the photon PDFs, we used the set by Aurenche, Fontannaz, and Guillet 
(AFG) \cite{AFG}.
In all cases, we found reasonable agreement between the experimental data and 
our NLO predictions as for both normalization and shape, as may be seen from
Fig.~\ref{fig:uni}.
We conclude that our global analysis of inclusive charged-hadron 
production provides evidence that both the predicted scaling violations and
the universality of the FFs are realized in nature.

\section{Inclusive $B$-Meson Production
\label{sec:four}}

The QCD-improved parton model implemented in the $\overline{\rm MS}$
renormalization and factorization scheme and endowed with nonperturbative
FFs, which proved itself so convincingly for light- \cite{bkk,kkp,kkp2} and
$D^{*\pm}$-meson \cite{dst} inclusive production, also provides an ideal
theoretical framework for a coherent global analysis of $B$-meson data
\cite{bme}, provided that $\mu\gg m_b$, where $\mu$ is the energy scale
characteristic for the respective production process.
Then, at LO (NLO), the dominant logarithmic terms, of the form
$\alpha_s^{n,n+1}\ln^n\left(\mu^2/m_b^2\right)$ with $n=1,2,\ldots$, are
properly resummed to all orders by the AP evolution, while power terms of the
form $\left(m_b^2/\mu^2\right)^n$ are negligibly small and can be safely
neglected.
The criterion $\mu\gg m_b$ is certainly satisfied for $e^+e^-$ annihilation on
the $Z$-boson resonance, and for hadroproduction of $B$ mesons with
$p_T\gg m_b$.
Furthermore, the universality of the FFs is guaranteed by the factorization
theorem \cite{col}, which entitles us to transfer information on how $b$
quarks hadronize to $B$ mesons in a well-defined quantitative way from
$e^+e^-$ annihilation, where the measurements are usually most precise
\cite{ob,ab}, to other kinds of experiments, such as hadroproduction
\cite{cdf1,cdf2}.

\begin{figure}[t]
\begin{center}
\begin{tabular}{ll}
\parbox{5.65cm}{
\epsfig{file=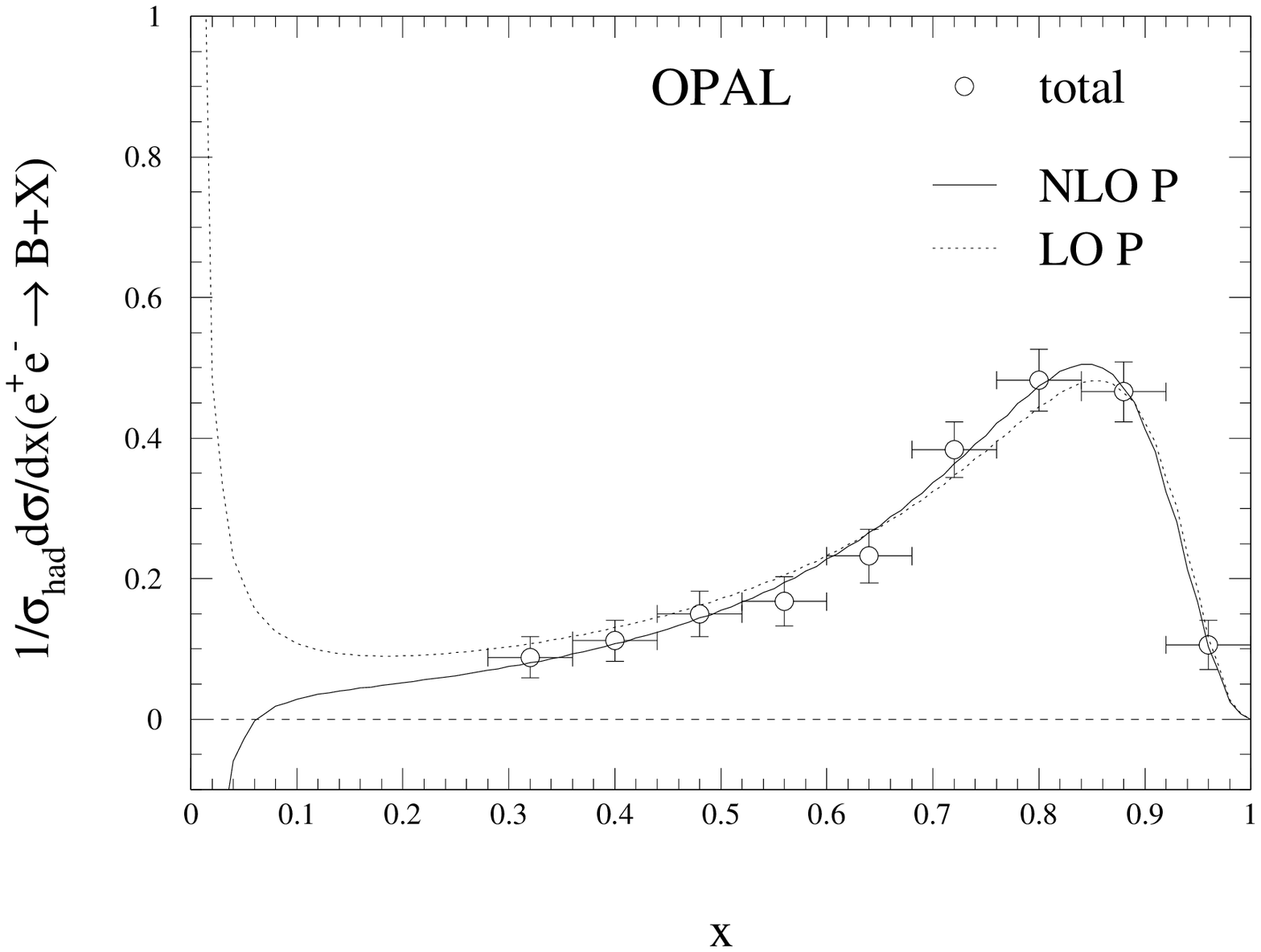,width=5.65cm,height=5.65cm,bbllx=9pt,bblly=21pt,%
bburx=510pt,bbury=397pt,clip=}
} &
\parbox{5.65cm}{
\epsfig{file=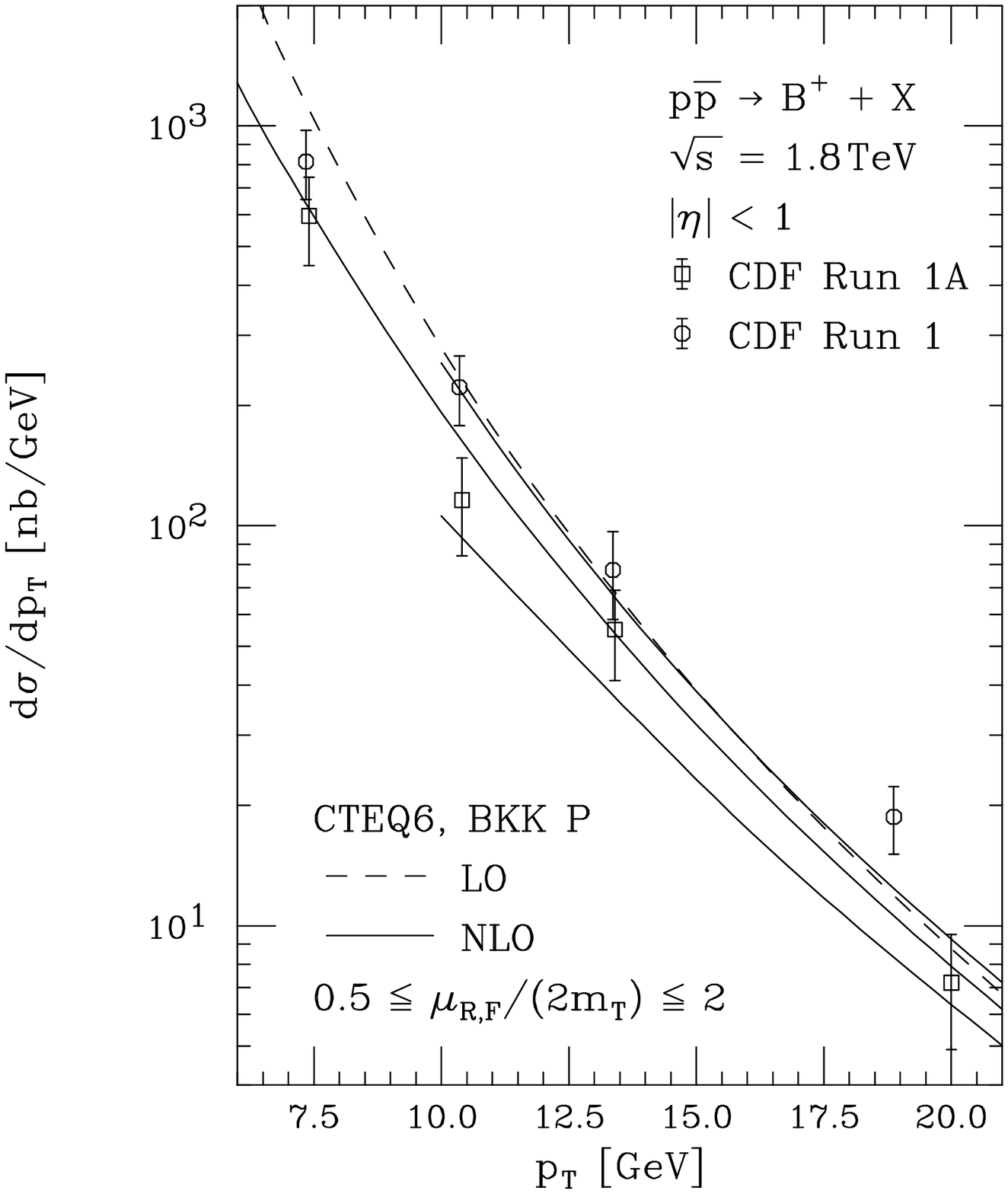,width=5.65cm,height=5.65cm,bbllx=52pt,bblly=102pt,%
bburx=532pt,bbury=671pt,clip=}
}
\vspace*{-0.3cm} \\
(a) & (b)
\end{tabular}
\caption[]{({\bf a}) Our LO and NLO fits \cite{bme} to OPAL data of
$e^+e^-\to B+X$ \cite{ob}.
({\bf b}) Comparison of CDF data of $p\overline{p}\to B+X$ \cite{cdf1,cdf2}
with our LO and NLO predictions \cite{bme}}
\label{fig:b}
\end{center}
\end{figure}

In \cite{bme}, the distribution in the scaled $B$-meson energy
$x=2E_B/\sqrt s$ measured by OPAL \cite{ob} at LEP1, which is compatible with
the subsequent measurements by ALEPH at LEP1 and SLD at SLC \cite{ab}, was
fitted at LO and NLO using three different ans\"atze for the $b\to B$ FF at
the starting scale $\mu_0=2m_b=10$~GeV.
The best fit was obtained for the ansatz by Peterson et al.\ (P) \cite{pet},
with $\chi_{\rm DF}^2=0.67$ (0.27) at LO (NLO) [see Fig.~\ref{fig:b}(a)].
The $\varepsilon$ parameter was found to be $\varepsilon=0.0126$ ($0.0198$).
We emphasize that the value of $\varepsilon$ carries no meaning by itself, but
it depends on the underlying theory for the description of the fragmentation
process $b\to B$, in particular, on the choice of the starting scale $\mu_0$,
on whether the analysis is performed in LO or NLO, and on how the final-state
collinear singularities are factorized in NLO.
In Fig.~\ref{fig:b}(b), the $B^+$-meson $p_T$ distribution measured by CDF in
Run~1A \cite{cdf1} and Run~1 \cite{cdf2} is compared with our LO and NLO
predictions evaluated using the CTEQ6 proton PDFs \cite{cteq6} and the
BKK-P $B$-meson FFs \cite{bme} and choosing $\mu_{R,F}=\xi_{R,F}\times2m_T$,
where $m_T=\sqrt{p_T^2+m_b^2}$.
The theoretical uncertainty at NLO is estimated by independently varying
$\xi_R$ and $\xi_F$ from 0.5 to 2 about the default value 1.
It only amounts to ${+21\atop-27}\%$ at $p_T=15$~GeV, but steadily increases
towards smaller values of $p_T$, reaching ${+33\atop-45}\%$ at $p_T=10$~GeV.
Variations in the proton PDFs and the ansatz for the $b\to B$ FF \cite{bme}
only reach a few percent.
We observe that the Run~1A data comfortably lies within the theoretical error
band, while the full Run~1 data tends to be somewhat on the high side,
especially in the upmost $p_T$ bin.

In the case of $\gamma\gamma\to D^{*\pm}+X$ at LEP2, the inclusion of
finite-$m_c$ effects was found to reduce the cross section by approximately
20\% (10\%) at $p_T=2m_c$ ($3m_c$) \cite{ks}, i.e., their magnitude is roughly
$m_c^2/p_T^2$, as na\"\i vely expected.
By analogy, one expects the finite-$m_b$ terms neglected in \cite{bme} to have
a moderate size, of order 20\% (10\%) at $p_T=10$~GeV (15~GeV).
This is considerably smaller than the scale uncertainty and appears
insignificant compared to the excess of the CDF data \cite{cdf1,cdf2} over the
traditional NLO analysis \cite{nas} in a scheme with $n_f=4$ active quark
flavours, where the $b$ quark is treated in the on-mass-shell renormalization
scheme.
In this massive scheme, there are no collinear singularities associated with
the outgoing $b$-quark lines that need to be subtracted and absorbed into the
FFs.
In fact, in this scheme there is no conceptual necessity for FFs at all, but
they are nevertheless introduced in an ad-hoc fashion in an attempt to match
the $B$-meson data.
However, in the absence of a subtraction procedure, there is also no
factorization theorem in operation to guarantee the universality of the FFs
\cite{col}.
By the same token, such FFs are not subject to AP evolution.
Thus, there is no theoretical justification to expect, e.g., that a single
value of the Peterson $\varepsilon$ parameter should be appropriate for
different types of experiment or at different energy scales in the same type of
experiment.
In other words, the feasibility of global data analyses is questionable in
this scheme.
Moreover, this scheme breaks down for $p_T\gg m_b$ because of would-be
collinear singularities of the form $\alpha_s\ln\left(p_T^2/m_b^2\right)$,
which are not resummed.

\begin{figure}[t]
\begin{center}
\begin{tabular}{ll}
\parbox{5.65cm}{
\epsfig{file=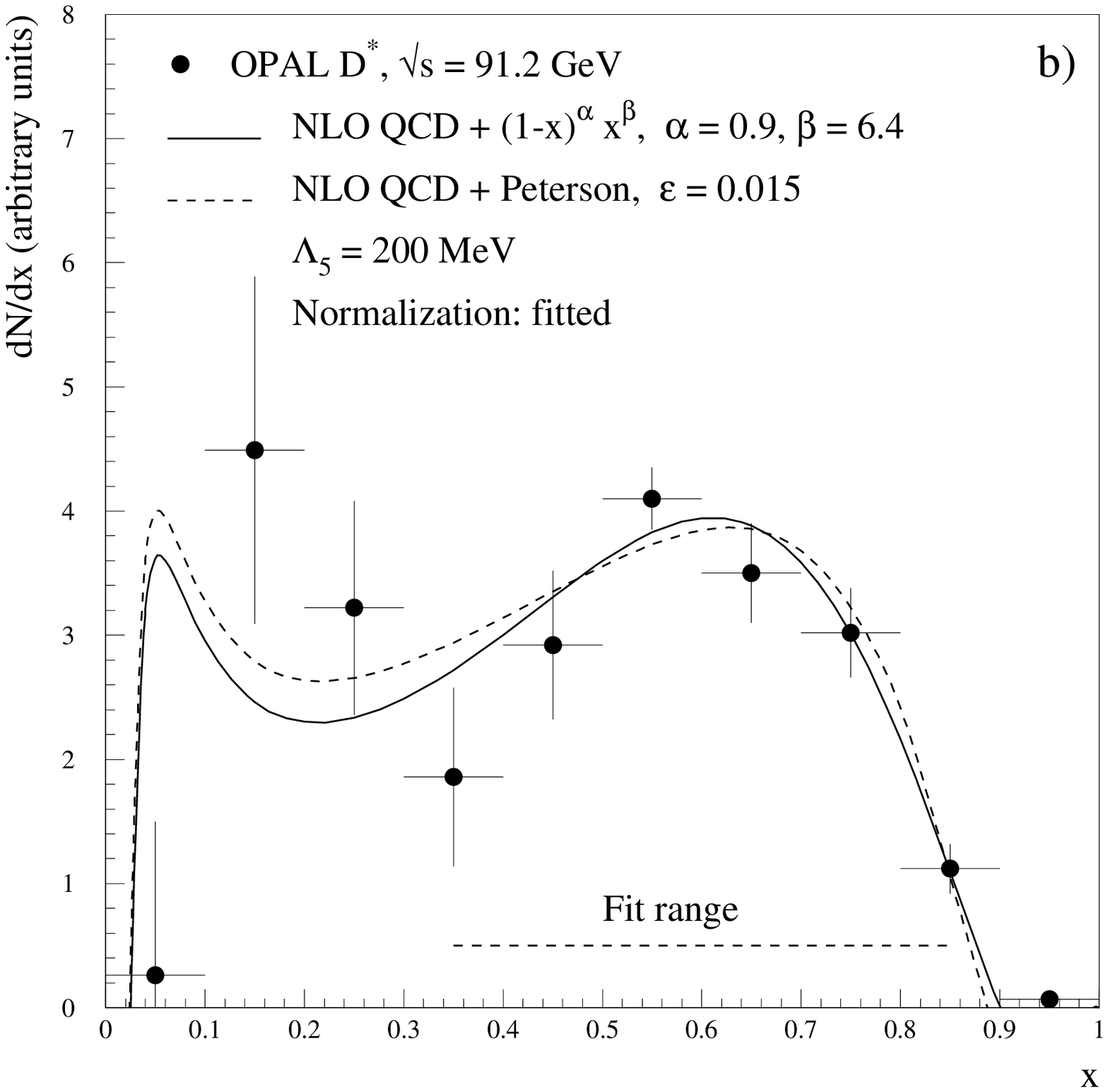,width=5.65cm,height=5.65cm,bbllx=28pt,bblly=161pt,%
bburx=523pt,bbury=650pt,clip=}
} &
\parbox{5.65cm}{
\epsfig{file=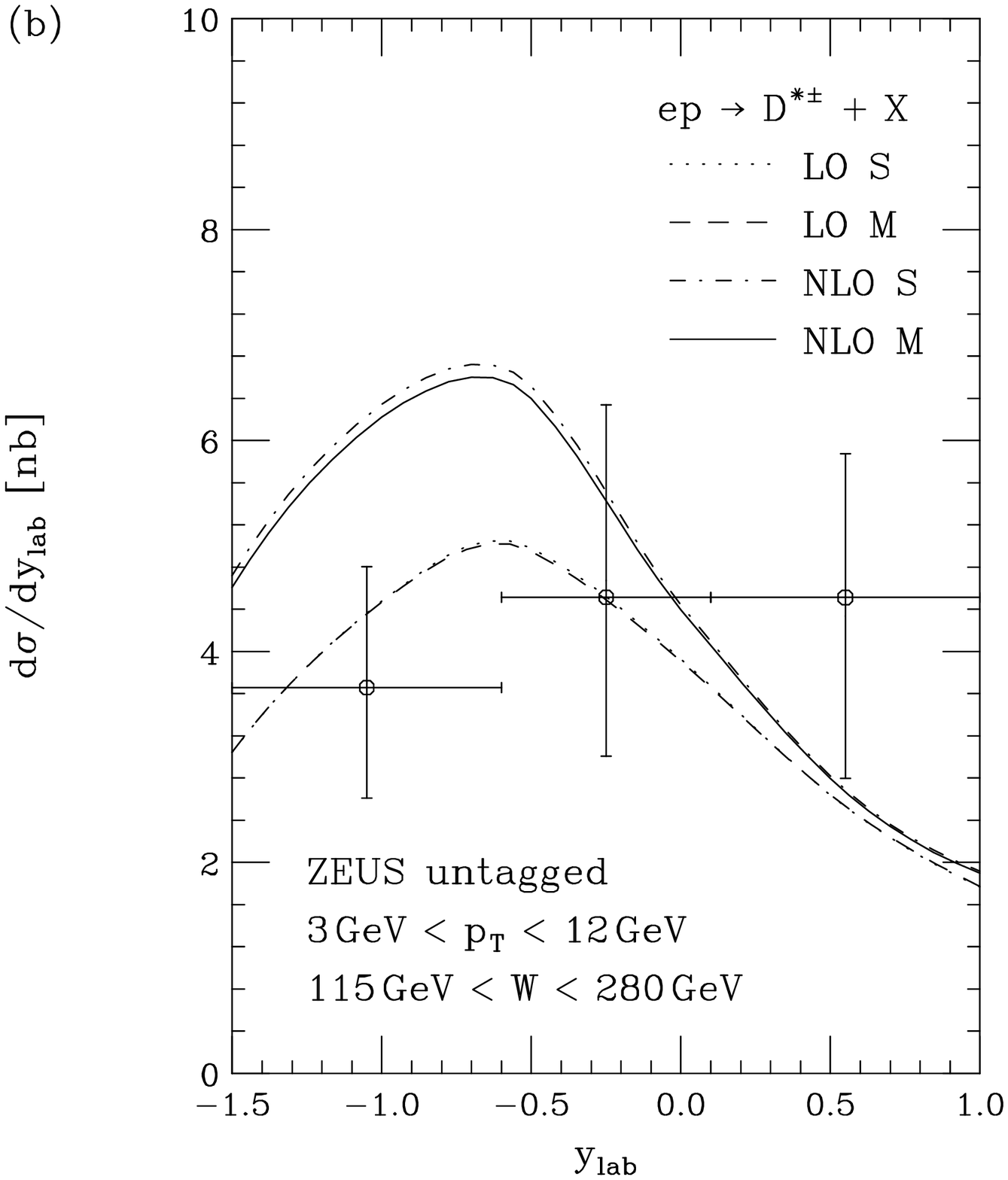,width=5.65cm,height=5.65cm,bbllx=52pt,bblly=102pt,%
bburx=543pt,bbury=677pt,clip=}
}
\vspace*{-0.3cm} \\
(a) & (b)
\end{tabular}
\caption[]{({\bf a}) NLO fit \cite{gre} to OPAL data of $e^+e^-\to D^{*\pm}+X$
using PFFs.
({\bf b}) Comparison of ZEUS data of $ep\to D^{*\pm}+X$ in photoproduction
with our LO and NLO predictions implemented with PFFs \cite{pff}}
\label{fig:pff}
\end{center}
\end{figure}

The attempt to split the $B$-meson FFs into a so-called perturbative FF (PFF)
and a nonperturbative remainder is interesting in its own right.
However, detailed analysis for $D^{*\pm}$-meson FFs \cite{pff} revealed that
such a procedure leads to deficient results in practical applications.
On the one hand, at NLO, the cross section $d\sigma/dx$ of $e^+e^-$
annihilation becomes negative in the upper $x$ range, at $x\agt0.9$
[see Fig.~\ref{fig:pff}(a)], where the data is very precise, so that a
low-quality fit is obtained unless this $x$ range is excluded by hand
\cite{pff,gre}.
On the other hand, the LO and NLO predictions for other types of processes,
such as photoproduction in $ep$ scattering [see Fig.~\ref{fig:pff}(b)],
significantly differ \cite{pff}, which implies that the perturbative stability
is insufficient.

The idea \cite{bme} of performing a coherent analysis of LEP1 and Tevatron
data of inclusive $B$-meson production was recently revived using an
unconventional scheme named FONLL, in which the traditional result in the
massive scheme \cite{nas} and a suitably subtracted result in a massless
scheme with PFFs are linearly combined \cite{cac}.
The degree of arbitrariness in this procedure may be assessed by noticing that
the massless-scheme term is weighted with an ad-hoc coefficient function of
the form $p_T^2/\left(p_T^2+25m_b^2\right)$ so as to effectuate its
suppression in the low-$p_T$ range and that this term is evaluated at
$p_T^\prime=\sqrt{p_T^2+m_b^2}$ while the massive-scheme term is evaluated at
$p_T$.
Since the FONLL scheme interpolates between the massive scheme and the
massless scheme with PFFs, it inherits the weaknesses of both schemes detailed
above.
In particular, the negativity of the NLO cross section of $e^+e^-\to B+X$ in
the upper $x$ range forces one to exclude the data points located there from
the fit.
In \cite{cac}, this is achieved by resorting to what is called there the
moments method, i.e., the large-$x$ region is manually faded out by selecting
one particular low moment of the $b\to B$ FF, namely the one corresponding to
the average $x$ value, thereby leaving the residual information encoded in the
data unused.

\section{Conclusions
\label{sec:five}}

We reviewed recent LO and NLO analyses of $\pi^\pm$, $K^\pm$, and
$p/\overline{p}$ FFs \cite{kkp}, which also yielded new values for
$\alpha_s^{(5)}(M_Z)$ \cite{kkp1}.
Although these FFs are genuinely nonperturbative objects, they possess two
important properties that follow from perturbative considerations within
the QCD-improved parton model and are amenable to experimental tests, namely
scaling violations and universality.
The scaling violations were tested \cite{kkp,kkp2} by making comparisons with
data of $e^+e^-$ annihilation at CM energies below \cite{low} and above
\cite{D2} those pertaining to the data that entered the fits.
The universality property was checked \cite{kkp2} by performing a global study
of high-energy data on hadroproduction in $p\overline{p}$ collisions
\cite{UA1,CDF} and on photoproduction in $e^\pm p$ \cite{H1} and $e^+e^-$
\cite{Ogg} collisions.
Our NLO FFs \cite{kkp} agree with other up-to-date sets \cite{kre} within the
present experimental errors \cite{kkp2}.

High-energy data on hadroproduction particularly probes the FFs in the
large-$x$ region, is especially sensitive to the gluon FFs, and, therefore,
carries valuable information complementary to that provided by $e^+e^-$ data.
This information, apart from being interesting in its own right, is
indispensible in order to reliably predict the $\pi^0$ background for the
$H\to\gamma\gamma$ signal of the intermediate-mass Higgs boson at the Tevatron
and the LHC.
It would thus be highly desirable if the experiments at the Tevatron made an
effort to update their analyses, which date back to 1988 \cite{CDF}.

Adopting a similar theoretical framework for $B$-meson production, we
demonstrated \cite{bme} that the notorious excess of the Tevatron data
\cite{cdf1,cdf2} over existing theoretical calculations can be ascribed, at
sufficiently large values of $p_T$, to nonperturbative fragmentation effects
inappropriately taken into account previously.
In combination with the factorization formalism of nonrelativistic QCD
\cite{bbl}, this framework also leads to a successful description \cite{kk} of
inclusive $J/\psi$ and $\psi^\prime$ production from $B$-meson decay
\cite{bpsi}.
A rigorous procedure to implement the finite-$m_b$ terms in an NLO framework
where terms of the form
$\alpha_s^{n+1}\ln^n\left(\mu^2/m_b^2\right)$ are resummed by AP evolution and
the universality of the FFs is guaranteed by the factorization theorem
\cite{col} is to directly subtract the would-be collinear singularities of the
form $\alpha_s\ln\left(p_T^2/m_b^2\right)$ in the massive-scheme result in a
way that conforms with the massless scheme \cite{ks} (see also \cite{sca}).

\vspace{1cm}
\noindent
{\bf Acknowledgements}
\smallskip

\noindent
The author is grateful to J. Binnewies, G. Kramer, and B. P\"otter for their
collaboration on the work presented here.
His research is supported in part by DFG Grant No.\ KN~365/1-1 and BMBF Grant
No.\ 05~HT1GUA/4.

\end{document}